\documentclass[authoryear,preprint]{elsarticle}
\usepackage{amsmath}
\usepackage{amssymb}
\usepackage{amsthm}
\usepackage{mathtools}
\usepackage{braket}
\usepackage{bm}
\usepackage{bbm}
\usepackage{graphicx}
\usepackage{xcolor}
\DeclarePairedDelimiter{\abs}{\lvert}{\rvert}
\newcommand{\numberset}{\mathbb}

\newcommand{\R}{\numberset{R}}
\newcommand{\Z}{\numberset{Z}}
\newcommand{\C}{\numberset{C}}

\newcommand{\sigv}{\sigma_\text{v}}

\newcommand{\hsigd}{{\sigma}_\text{d}}
\newcommand{\bx}{\mathbf{x}}
\newcommand{\ri}{i}
\newcommand{\bC}{\mathbf{C}}
\newcommand{\bCo}{\mathbf{\bar{C}}}
\newcommand{\be}{\mathbf{e}}
\newcommand{\bE}{\mathbf{E}}
\newcommand{\bF}{\mathbf{F}}

\newcommand{\bG}{\mathbf{G}}
\newcommand{\bH}{\mathbf{H}}
\newcommand{\bI}{\mathbf{I}}
\newcommand{\bbm}{\mathbf{m}}

\newcommand{\bX}{\mathbf{X}}

\newcommand{\bna}{\boldsymbol{\nabla}}

\begin{document}

\title{Crystal elasto-plasticity on the Poincar\'e half-plane}

\date{\today}
\author[1]{Edoardo Arbib}
\ead{edoardo.arbib@mail.polimi.it}
\address[1]{Department of Physics, Politecnico di Milano, Italy}
\author[1]{Paolo Biscari}
\ead{paolo.biscari@polimi.it}
\author[3]{Luca Bortoloni}
\address[3]{LS `Galileo Galilei', Alessandria, Italy}
\ead{luca.bortoloni@istruzione.it}
\author[4]{Clara Patriarca}
\ead{clara.patriarca@polimi.it}
\address[4]{Department of Mathematics, Politecnico di Milano, Italy}
\author[5]{Giovanni~Zanzotto}
\ead{giovanni.zanzotto@unipd.it}
\address[5]{DPG, Universit\`a di Padova, Italy}
\fntext[6]{We acknowledge the contribution of Dr.\ Marc Lesimple in the early stages of this work, and the financial support from the Italian PRIN project 2017KL4EF3. G.Z. thanks Dr.\ Gareth Parry for several conversations.}

\begin{abstract}

We explore the nonlinear variational modelling of two-dimensional (2D) crystal plasticity based on strain energies which are invariant under the full symmetry group of 2D lattices. We use a natural parameterization of strain space via the upper complex Poincar\'e half-plane. This transparently displays the constraints imposed by lattice symmetry on the energy landscape.
Quasi-static energy minimization naturally induces bursty plastic flow and shape change in the crystal due to the underlying coordinated basin-hopping local strain activity.
This is mediated by the nucleation, interaction, and annihilation of lattice defects occurring with no need for auxiliary hypotheses. Numerical simulations highlight the marked effect of symmetry on all these processes.  The kinematical atlas induced by symmetry on strain space elucidates how the arrangement of the energy extremals and the possible bifurcations of the strain-jump paths affect the plastification mechanisms and defect-pattern complexity in the lattice.
\end{abstract}


\maketitle

\section{Introduction}

Crystal elasto-plasticity, especially at the microscales, represents a peculiar meeting point between coexisting effects such as solids elasticity and plastic flow with dislocation-driven intermittency, patterning, and hardening, see \cite{zapperi02}, \cite{dislocationstructures1}, \cite{dimiduk06}, \cite{zapperisciencebursts}, \cite{fressengeas09}, \cite{uchic092}, \cite{dislocationstructures2}, \cite{papanikolau}, \cite{hardening}. Within the framework of conti\-nuum mechanics, linear elasto-plasticity satisfactorily explains most of crystals' behavior at the macroscopic scales. As the elastic and plastic distortions are largely independent at these scales, a number of ad-hoc modelling assumptions need to be introduced, including yield conditions and plastic flow rules \citep{gurtin, bookplasticitylee}.

At the microscopic scales, however, elasto-plastic effects are inherently coupled, and there has been extensive research on unified modeling approaches. Phase-field models have been successful in reproducing the evolution of the ground-state configurations which is at the basis of crystal plasticity, especially in the more recent nonlinear implementations by \cite{levitas1, levitas2} or \cite{biurzaza}, \cite{denoualJMPLplasticity}, \cite{kan}. Atomistic and discrete dislocation dynamics (DDD) models, as in
\cite{phasefieldcrystaldahmen}, \cite{zapperiDDDreview}, \cite{acklandasymmetry}, \cite{xiang}, \cite{zalezak}, \cite{plastMDbulatovnature}, \cite{hardening}, have also greatly expanded our knowledge of crystal microplasticity phenomena.

However, on the one hand both phase-field and DDD methods need auxiliary hypotheses on  plastic flow and/or defect nucleation and interaction. Atomistic-type modelling, on the other hand, while free from such drawbacks, concerns only short time scales, and is also not efficient in elucidating how crystal symmetry and kinematics influence defect and microstructure formation and evolution in the distorted solids. For instance, the fundamental role of kinematic compatibility in the mechanical behavior of crystalline materials may be appreciated only through an in-depth analysis of the deformation-gradient map representing the lattice distortion \citep{13james_nat, biurzaza, jamesreview, james2D}.

To sidestep the above-mentioned issues in the modelling of crystal mechanics, in this paper we pursue an approach to crystal elasto-plasticity based on a nonlinear continuum-type variational framework explicitly informed by global crystal symmetry. The behavior of the deforming material is understood in terms of energy-minimizing strain fields on a suitable energy landscape, whose topography is determined by the full invariance of the underlying lattice. This makes non-convex the strain energy density, with a symmetry-prescribed countable infinity of minimizers. The associated valley floors \citep{valleyfloor1} give the preferential avenues on which occurs the coordinated basin-hopping of the local strains, strongly directing the progress of the body's deformation field on the bumpy landscape of the total energy functional. In our lattice-based, and thus discrete, systems such evolution is regularized \citep{PRLgruppone}, and bursty material response emerges as a discrete sequence of local minimizers is visited under the external driving. In this context crystal plasticity results from the formation and propagation of crystal defects, such as dislocations, with no need for extra assumptions on their nucleation, annihilation and movement, as these derive from concentrated slip processes which allow large strain relaxation in the crystal while seeking to locally preserve the lattice structure.

Inertia and the possible presence of oscillatory external forces may influence the bursty dynamics of crystals, for instance affecting the scaling behaviour and shape of plastic avalanches \citep{truvain08, salrob13, weiss19, dahmen19}. Such effects may become relevant or even dominate in particular situations, especially when the characteristic time of the driving forces approaches the avalanche durations \citep{dahmen19}, or
when the dislocations are induced to move at high velocities \citep{tang18}. However, when the driving and the relaxation time scales are
sufficiently separated, these effects may be neglected, and it is possible to determine the equilibria through local energy-minimization as a
limit case of over-damped dynamics driving the system towards some nearest optimal configurations on a very slowly varying energy
landscape. Also the viscous contributions that may in general affect the way and the characteristic times in which local equilibrium is approached, may be expected to have a less significant role in these quasi-static processes where the system visits a sequence of equilibria under the very slow driving. The role of time-scale separation and the possibility of using energy minimization for the slower relaxation phenomena have been discussed for instance in relation to slip concentration by \cite{freddi16}, or in shear band propagation in the presence of fracture by \cite{arriaga17}.

In more detail, we model crystal elasto-plastic behavior based on the original proposal by \cite{eri77, eri80} concerning the material symmetry of a crystalline substance. Accordingly,  we assume the stain energy density $\sigma$ to be invariant under all the deformations which map the underlying lattice onto itself, see also \cite{folkins}, \cite{parry}. These lattice-invariant distortions thus dictate, in the space of strain tensors, the location of infinitely-many relaxed states for the crystal. Explicitly, these are given by the elements of (a conjugate of) the infinite and discrete group $\text{GL}(n,\Z)$ collecting all the invertible $n \times n$ matrices with integral entries, where $n=2,3,$ is the dimension of the lattice under study. For brevity, we refer to this property as GL-invariance.

Extensive research has successfully used the ensuing variational models. Special attention has been given to reversible martensitic transformations, with the aim of better understanding and enhancing the performance of shape-memory alloys
\citep{PZbook, bhattabook, jamesperspective1, 13james_nat, jamesperspective2, jamesreview}. In these cases the associated finite deformations are largely confined to suitable `Ericksen-Pitteri neighborhoods' (EPNs) in strain space, whereon GL-symmetry reduces to point-group symmetry \citep{eri80, pitterireconciliation, PZbook, ContiZanzotto}. This much reduces the difficulties related to the full GL-invariance of the theory. GL-symmetry has also been used to study reconstructive structural transformations, in which strains may reach or overcome the EPN bundaries, producing lattice-defects and plasticity phenomena \citep{PZbook, ContiZanzotto, BCZZnature, ironshear1, ironshear2, pacoreview, corridoifrancesi, denoualJMPLplasticity, pacoreview2, pacoreview3}. Full lattice invariance has furthermore been used for instance in the study of slip and twinning in helical structures \citep{james2D}. In general, GL-invariance plays a key role in the determination of lattice twinning mechanisms \citep{PZbook,
bhattabook}, in relation to both structural transformations and plasticity in crystals. Depending on circumstance, dislocation-based irreversible
plastic phenomena may be assisted also by multiple-cell twinning deformations, related to the fact that crystals can de-couple internal
degrees of freedom behaving as ‘multi-lattices’ (unions of congruent Bravais lattices), rather than always as a single Bravais lattice. The details of
the more complex kinematics and energetics needed for the extension of the present GL-approach to encompass also these deformation modes can be found in \cite{PZbook}, see also \cite{jamesreview}. \cite{prolifarationtwinning} give the corresponding treatment of twinning proliferation in magnesium, while \cite{ironshear2} use this framework to establish a bridge between GL-invariant elasto-plasticity and quantum mechanical approaches to strain-energy computations. Thus, the proposed line of research based on GL-energetics provides an attractive avenue also for the investigation of actual crystal plasticity. Besides the mentioned benefits afforded by the continuum approach, a further main point of interest here is the possibility of capturing in an intrinsic way the differences reported in experiments on the plastic behavior of crystals with different symmetry, orientation, or loading \citep{greer, uchic092, differencesymmetry1, differencesymmetry2, papanikolau, differenceloading, nonschmid2, nonschmid1, sparks1, sparks2}.

In the present work we advance this theoretical approach, evidencing further basic effects GL-elasto-plasticity in 2D. While interesting \emph{per se}, this also provides a necessary road-map for the much more complex study of the 3D case. For recent results on 3D crystal plasticity via phase-field modelling under full lattice symmetry, see \cite{biurzaza}.

As in \cite{folkins}, \cite{parry}, \cite{PRLgruppone}, here we use a natural parameterization of 2D strain space by means of the upper complex Poincar\'e half-plane $\mathbb{H}$, one of the best known models of the hyperbolic plane \citep{poincarehalfplane2, poincarehalfplane1}. On $\mathbb{H}$ the well-known Dedekind tessellation \citep{dedekind1, dedekind5} transparently displays the action of GL-symmetry on strain space, and thus on a crystal's energy landscape. By elaborating on the ideas by \cite{parry}, we write explicitly 2D-lattice strain energies which automatically comply with GL-invariance by using suitable modular forms, a well known class of complex functions considered in various branches of mathematics and physics \citep{modularforms2, modularforms1, dedekind3, dedekind1, modularphysics}. By design, in the present study of crystal plasticity we restrict ourselves to GL-energies with a minimum complexity of the associated topography. We thus consider a family of potentials with a \emph{single} minimizer (periodically replicated through GL-periodicity) and a very simple explicit expression, see also \cite{PRLgruppone}. This is motivated by the aim of obtaining plastic-flow results without possibly spurious effects due to the presence of other metastable lattice configurations. Similar methods can be used to write also GL-invariant potentials with two or more minimizers coexisting in the fundamental domain of GL-periodicity (see \cite{tesipatriarca} for preliminary results), thus allowing for the modelling of crystal plasticity and structural phase transformations, as in \cite{ContiZanzotto}.

The performed numerical simulations highlight the influence of GL-con\-strained energy topography on plastification and defect-generation mechanisms. In particular it is evidenced the key role played by the symmetry-constrained arrangement of the ground states and of the possible bifurcations of the strain-jump paths.

This paper is organized as follows. In Section~2 we give preliminaries on the GL-invariant energies for crystalline materials. In Section~3 we present the Poincar\'e half-plane $\mathbb{H}$ as a parameterization of strain space in 2D, and we introduce the Klein invariant $J$, by means of which we construct a class of simple GL-invariant energies in Section~4. We use the latter for the numerical simulations of crystal plasticity presented and discussed in Section~5. In Section~6 we stress the role on the plastification mechanisms of the GL-constrained networks built by means of the energy extremals and valley floors.

\section{Strain energies for crystalline materials}

In a two-dimensional (2D) setting, we consider a body whose points $\bX$ have coordinates $X_i$ ($i=1,2$) with respect to an orthonormal basis. The deformations are one-to-one maps $\bx = \bx(\bX)$, where the $x_i$ ($i=1,2$) identify current point positions. The deformation gradient $\bF=\bna\bx$ is a $2\times 2$ matrix with elements $F_{ij}=\partial x_i/\partial X_j$, and the associated symmetric positive-definite Cauchy-Green tensor is
\begin{equation}
\bC=\bF^T\bF = \bC^T>0, \quad \quad \bC=2\bE+\bI,
\label{eq:definitionstrain}
\end{equation}
where $\bE$ is the nonlinear strain \citep{PZbook, gurtin}. The unimodular tensors
\begin{equation}
\bCo=(\det\bC)^{-1/2}\bC,
\label{unimodular}
\end{equation}
with $\det \bCo =1$, describe a 2D hyperboloid within the 3-dimensional cone of $2\times 2$ symmetric positive-definite tensors $\bC$ in (\ref{eq:definitionstrain}).

The total strain-energy functional is the integral of a strain-energy density over the reference configuration of the body. The energy density $\sigma$ is a smooth Galilean-invariant scalar function, which depends on $\bF$ only through $\bC$, i.e.
\begin{equation}
\sigma= \sigma(\bC),
\label{strainenergy}
\end{equation}
with material symmetry requiring that
\begin{equation}
\sigma(\bC)=\sigma(\bG^T\bC\bG),
\label{eq:energyinvariance}
\end{equation}
for any $\bC = \bC^T>0$ and any deformation $\bG \in \cal G$. Here  $\cal G$ is a suitable group contributing to the characterization of the material response. As mentioned in the Introduction, for 2D crystalline substances we follow Ericksen's earlier proposal and assume $\cal G$ to be conjugate, via the choice of reference basis, to the group $\text{GL}(2,\Z)$ describing the global symmetry of 2D Bravais lattices
\citep{eri80, michel, PZbook}.

\begin{figure}

\centering

\includegraphics[height=4.8cm, width=8.5cm]{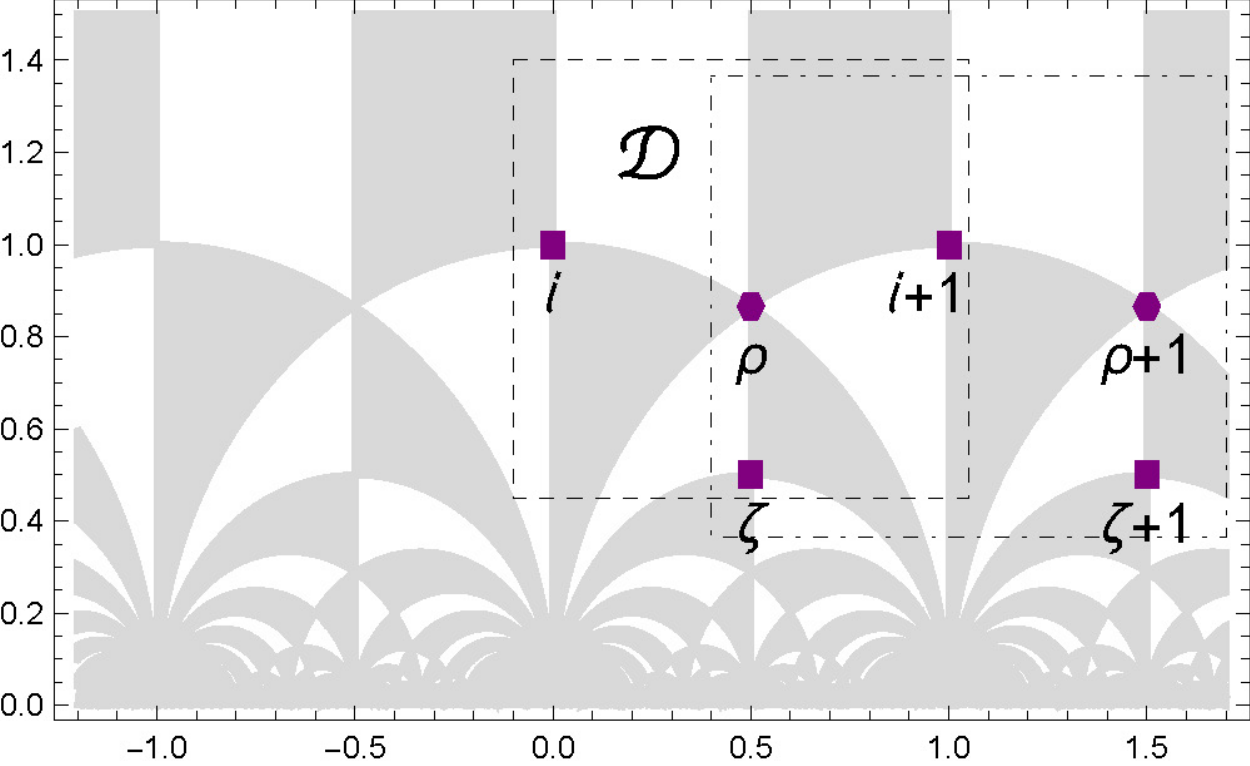}

\caption{(Color online) The Dedekind tessellation of the Poincar\'e upper complex half-plane $\mathbb{H}$. This gives a global atlas of the action of $\text{GL}(2,\Z)$-symmetry on $\mathbb{H}$, and thus, via Eq.~(\ref{bridge}), on the space of unimodular strain tensors for 2D lattices. The standard fundamental domain $\cal D$ in Eq.~(\ref{fundomain}) is indicated. The interior[boundary] points of $\cal D$ correspond to lattices with [non-]trivial symmetry. Four equivalent square points  \big($i$, $i + 1$, $\zeta= \frac{1}{2}(i+1)$, $\zeta+1$\big),
and two equivalent hexagonal points \big($\rho = e^{i \pi/3} = \frac{1}{2}+\frac{\sqrt{3}}{2}i$, $\rho +1$\big)
are indicated. The highlighted windows are domains on $\mathbb{H}$ pertaining to the two numerical simulations of plastic flow presented in Section 5 (dashed window: square simulation, see Fig.~2; dot-dashed window: hexagonal simulation, see Fig.~3). See also the two Supplementary Videos.}
\label{fig:Dedekind}
\end{figure}

The strain energy is split into a convex volumetric part $\sigv$ penalizing the departure of $\det\bC$ from 1, plus a distortive term $\hsigd$ which only depends on $\bCo$:
\begin{equation}
 \sigma(\bC)=\sigv(\det\bC)+\hsigd(\bCo).
\label{eq:sigelastic}
\end{equation}
By GL-invariance, $\hsigd$ is non-convex with countably-many local minimizers, whose arrangement in strain space we discuss below (see Fig.~\ref{fig:Dedekind}).

\section{Parameterization of 2D distortions on the Poincar\'e half-plane}

The Poincar\'e (upper complex) half-plane is the set $\mathbb{H} = \{ x+\ri y\in\C$, $y>0\}$, endowed with the metric $(ds)^2=\big((dx)^2+(dy)^2\big)/y^2$, see \cite{poincarehalfplane2, poincarehalfplane1}. A complex representation of the cone of tensors $\bC = \bC^T>0$ is provided by associating to any $\bC$ the complex number as in \cite{folkins}, \cite{parry}:
\begin{equation}
\label{bridgelarge}
\hat z(\bC)=\frac{C_{12}}{C_{11}} + i\,\frac{\sqrt{\det{\bC}}}{C_{11}}
\in \mathbb{H}\, ,
\end{equation}
where $C_{ij}$ are the matrix elements of $\bC$ in the given orthonormal basis.
From Eq.~\eqref{bridgelarge} we obtain $\hat z(\bC)= \hat z(\bC') $ if and only if $ \bC$ and $\bC'$ are proportional. This means that the 2D $\bCo$-hyperboloid of unimodular strains in Eq.~(\ref{unimodular}) is smoothly mapped one-to-one onto $\mathbb{H}$ by:
\begin{equation}
\label{bridge}
\hat z(\bCo)={\bar{C}}_{11}^{-1}({\bar{C}}_{12} + i).
\end{equation}
The local body deformation, described by the four entries $F_{ij}$ of the deformation gradient $\bF$, is thus equivalently accounted for by using $\det\bF$, the angle $\theta$ in the polar decomposition of $\bF$, plus two independent variables in $\bCo$ (or the corresponding $\hat{z}(\bCo)$ from Eq.~(\ref{bridge})) which enter $\hsigd$ in Eq.~(\ref{eq:sigelastic}).

The identification in Eq.~(\ref{bridge}) is very useful in the present approach to crystal mechanics because the material-symmetry maps $\bC \mapsto \bG^T\bC\bG$, for  $\bG \in \cal G$, correspond via (\ref{bridge}) to the natural action on $\mathbb{H}$ by the classical fractional linear (Moebius) isometries of $\mathbb{H}$ with integral entries, augmented by the map $z \mapsto -\bar z$, where $\bar z$ is the conjugate to $z \in \C$ \citep{folkins, parry}. Explicitly, if $\bbm\in\text{GL}(2,\Z)$ and $\tilde{\bC}=\bbm^T\bC\bbm$, the corresponding map $z\mapsto \tilde z$ on $\mathbb{H}$ via~\eqref{bridge} is given by
\begin{align}
\tilde z=\begin{cases}
\displaystyle \frac{m_{22}z+m_{12}}{m_{21}z+m_{11}} &\text{if }\det \bbm=1 \\
\noalign{\bigskip}
\displaystyle \frac{-m_{22}\bar z+m_{12}}{-m_{21}\bar z+m_{11}}  &\text{if }\det \bbm=-1.
\end{cases}
\label{action}
\end{align}
A very effective geometrical representation of this action is given by the $\text{GL}(2,\Z)$-generated Dedekind tessellation of $\mathbb{H}$ shown in Fig.~\ref{fig:Dedekind} \citep{dedekind1, dedekind5}. It contains all the $\text{GL}(2,\Z)$-related mutually congruent (in the sense of the hyperbolic metric) copies of the fundamental domain
\begin{equation}
\label{fundomain}
\mathcal{D}=\{ z\in \mathbb{H} : \abs{z}\ge 1,\ 0\le\text{Re}(z)\le\tfrac{1}{2} \}.
\end{equation}
The points in the interior of $\mathcal{D}$ correspond through Eq.~(\ref{bridge}) to strain tensors (and thus lattices) with trivial symmetry and thus to oblique lattices, while the points on the boundary $\partial\mathcal{D}$ of $\mathcal{D}$ correspond to metrics possessing nontrivial symmetries. The latter include (see Fig.~\ref{fig:Dedekind}) rectangular lattices, represented by boundary points on the imaginary axis; fat-rhombic lattices (whose inner angles are greater than $\frac{\pi}{3}$), represented by boundary points with $\abs{z}=1$; and skinny-rhombic lattices, represented by boundary points with $\text{Re}(z)=1/2$. We refer to \cite{parry}, \cite{michel}, \cite{PZbook}, \cite{ContiZanzotto} for a detailed description of the global and point symmetry of the Bravais lattice types in 2D, and of the $\text{GL}(2,\Z)$ fundamental domain in $\bCo$-space.

The identification (\ref{bridge}) of the $\bCo$-hyperboloid with $\mathbb{H}$ leads to the remarkable observation by \cite{folkins}, \cite{parry}, that the GL-invariant strain energies $\hsigd$ for 2D crystalline materials in (\ref{eq:energyinvariance})-(\ref{eq:sigelastic}) are closely related to the modular functions on $\mathbb{H}$. Particularly useful for us is the existence \citep{modularforms2, modularforms1} of a unique complex function $J$, also known as the Klein invariant, holomorphic on $\mathbb{H}$ with the following properties: $J$ is periodic under the action (\ref{action})$_1$ of the 'modular group' $\text{SL}(2,\Z)$ (which denotes the subgroup collecting the positive-determinant elements of $\text{GL}(2,\Z)$, see \cite{dedekind5}); $J$  is one-to-one between the fundamental domain $\cal D$ and the complex half-plane $\C^+$ with non-negative imaginary part; furthermore, $J$ satisfies $J(z)=\overline{J(-\bar{z})}$ for any $z$, it assumes values $J(i) = 1$ and $J(\rho) = 0$, and it diverges when $\text{Im\,} z\to+\infty$, with a simple pole at infinity. Its only stationary points are  $z = i$ and $ z =\rho$  (i.e.~$J'(i) = 0$ and $J'(\rho) = 0$), with also $J''(\rho) = 0$.
The symmetry requirements imply that $J$ assumes real values at the boundary $\partial \cal D$ of $\cal D$, and that $J$ diverges also when $z$ approaches any rational point on the real axis, see also \cite{modularforms2}, \cite{modularforms1}, \cite{parry}, and Eq.~(\ref{furierj}) below.

\section{A class of simplest GL-invariant strain-energies for 2D crystal elasto-plasticity}

Based on the strain parameterization (\ref{bridge}) and the above properties of $J(z)$, \cite{parry} proposed that potentials satisfying (\ref{eq:energyinvariance})-(\ref{eq:sigelastic}) be written as suitable functions of $J$:
\begin{equation}
\label{energyJ}
\hsigd(\bCo) =\hsigd\big(J(\hat{z}(\bCo)) \big).
\end{equation}
As mentioned in the Introduction, depending on the specific properties and number of minimizers and other extremals in the domain $\cal D$ of Eq.~(\ref{fundomain}), energies of this form can describe plasticity and phase transformations in 2D crystals, as well as their coupling. Our present aim is deriving a family of simplest potentials in (\ref{eq:energyinvariance})-(\ref{eq:sigelastic})-(\ref{energyJ})  by using functions $\hsigd(J)$ in \eqref{energyJ}  which have a \emph{single} minimizer $z_0$ in $\cal D$, see also \cite{PRLgruppone}. A single minimizer, GL-related to $z_0$, is therefore also on each tile in the Dedekind tessellation of Fig.~\ref{fig:Dedekind}.

Given any energy-minimizing $\bCo_0$, we set $z_0 = \hat{z}(\bCo_0)$. Then $J_0(z) = |J(z) - J(z_0)|$ provides a basic measure of the distance between any strain and the ground state, via their counterparts $z\in\mathbb{H}$. By considering in (\ref{energyJ}) functions
\begin{equation}
\label{energyJ0}
\hsigd = \hsigd(J_0(z)) , \quad J_0(z) = |J(z) - J(z_0)|,
\end{equation}
it is possible to obtain a simple class of smooth potentials whose sole minimizers are $\bCo_0$ and all its GL-related copies when $z_0\in\partial\mathcal{D}$\footnote{As remarked in \cite{modularforms2}, \cite{modularforms1},
$J(z)$ is SL-invariant under (\ref{action})$_1$. However, a function $\hsigd(J_0(z))$ in (\ref{energyJ0}) is in fact GL-invariant
whenever the minimizer $z_0 = \hat{z}(\bCo_0)$ is on the boundary of $\cal D$, as in this case $J(z_0)$ is real. Ground states lying on $\partial\cal D$ correspond to lattices with non-trivial symmetries as in square, hexagonal, rhombic, or rectangular 2D Bravais lattices. Indeed, in order to have GL-invariance, strain functions as in (\ref{energyJ0}) must depend on (Im\,$J(z))^2$, while they can depend arbitrarily on Re\,$J(z)$. If $z_0$ has non-trivial symmetry, any function of $J_0(z)$ automatically depends on (Im\,$J(z))^2$, and does exhibit full GL-symmetry. If $z_0$ belongs to the interior of $\cal D$, thus corresponding to an oblique 2D-lattice ground state, potentials (\ref{energyJ0}) depending on $J_0$ would be SL- but not GL-invariant. In this case we can instead consider in (\ref{energyJ0}) functions for instance of $J_0^*(z)=\big([\text{Re}\,J(z)-\text{Re}\,J(z_0)]^2+ [(\text{Im}\,J(z))^2-(\text{Im}\,J(z_0))^2]^2\big)^{1/2}$}, which gives an alternative GL-invariant definition of the distance from the ground state.
To agree with standard linear elasticity when the strained lattices are sufficiently close to the well bottoms, the Taylor expansion of the functions in (\ref{energyJ0}) must have positive-definite quadratic behavior close to $z_0$, i.e.~$\sigma_{\text{d}}(J_0(z) ) \approx (z-z_0)^2$ for $z\approx z_0 \in \mathbb{H}$. Since $J_0(z)\approx |J'(z_0)| |z-z_0|$ as $z\approx z_0$, when $J'(z_0)\neq 0$ the simplest potential satisfying our requirements is
\begin{equation}
\hsigd\big(\bCo\big)= \mu J_0(z)^2= \mu |J(z) - J(z_0)|^2\quad \text{if} \quad J'(z_0)\neq 0,
\label{eq:nonsing}
\end{equation}
where $\mu >0$ is an elastic modulus, and $z = \hat{z}(\bCo)$. As $J'(z_0)\neq 0$ for all $z_0$ different from the corner points $i$ or $\rho$ of $\cal D$ in Fig.~\ref{fig:Dedekind}, the potential in (\ref{eq:nonsing}) is thus the desired one for all $z_0 \in\partial \cal D$, $z_0 \neq i, \rho$.

The particularly relevant cases $z_0=i$ (square) and $z_0=\rho$ (hexagonal), which represent the two maximally-symmetric lattices in 2D, need care because they are the only stationary points of $J$, so (\ref{eq:nonsing})$_2$ does not hold. The Klein invariant indeed satisfies $J(z)=1+O(z-i)^2$ as $z\to i$, and $J(z)=O(z-\rho)^3$ as $z\to\rho$,
%because $J'(i) = 0$ and $J'(\rho) = J''(\rho) = 0$, see \cite{modularforms2}, \cite{modularforms1}
because $J'(i) = 0 = J'(\rho) = J''(\rho) = 0$. To ensure a correct linear-elastic behavior we must then consider suitably modified potentials in (\ref{energyJ0})-(\ref{eq:nonsing}), i.e. $(|J(z)-J(i)|^{1/2})^2=|J(z)-1|$ for square lattices ($z_0=i$), and $(|J(z)-J(\rho)|^{1/3})^2=|J(z)|^{2/3}$ for hexagonal lattices ($z_0= \rho$) .

Summarizing, a unified expression for the simplest elasto-plastic GL-potentials for 2D crystals with square, hexagonal, rhombic or rectangular ground state $z_0 = \hat{z}(\bCo_0) \in\partial\mathcal{D}$ is
\begin{equation}
\label{potz0}
\sigma_{\text{d,plast}}\big(\bCo\big) = \mu J_0(z)^{2/\kappa(z_0)}= \mu \big|J(z)-J(z_0)\big|^{2/\kappa(z_0)}, \quad z = \hat{z}(\bCo),
\end{equation}
where $\mu >0$ and $\kappa(z_0)$ is the order of zero of $J(z)-J(z_0)$ near $z_0$ (in particular, $\kappa(z_0)=1$ if $J'(z_0)\neq 0$, i.e.~for all $z_0 \in \partial \mathcal{D}\setminus \{ i,\rho\}$).

We notice that writing $\hsigd$ in (\ref{potz0}) in terms of the above $J$-based modular order parameters further develops the notion of transcendental order parameters considered in the extended Landau theory for crystalline materials examined in \cite{transcendental1}, \cite{transcendental2}, as $J_0(z)$ endows the strain energy density $\sigma$ in (\ref{eq:sigelastic}) with the full invariance of the underlying (2D) lattice. The strain-energies $J_0(z)$ in (\ref{potz0}) are a simplest class of fully GL-invariant regular functions possessing a unique minimizer $z_0$
in the fundamental domain $\cal D$. Given this, $J_0(z)$ then has in $\cal D$ at most three stationary points, i.e.~the imposed minimizer $z_0$, and the two further points $i$ and $\rho$ where any regular GL-invariant function, in particular as in (\ref{potz0}), is necessarily stationary due to GL-symmetry. The two cases where either $z_0 = i$ (energy with square ground states) or $z_0 = \rho$ (energy with hexagonal ground states) are explicitly considered below. Besides providing simple strain-energies exhibiting the full lattice symmetry and a tunable ground state, the functions (\ref{potz0}) may be used as building blocks within the wider family of energies (\ref{energyJ}). They can for instance provide explicit expressions, both analytic and GL-invariant, for potentials in structural phase transformations and their associated elasto-plastic effects, as in \cite{tesipatriarca}. They can also be used to describe the generalized stacking fault energies ($\gamma$-surfaces) derived through ab-initio \citep{bakst18,kamimura18} or atomistic phase field \citep{qiu19} computations. Polynomial combinations of $\sigma_{\text{d,plast}}$ in (\ref{potz0}) give the simplest class of functions for these purposes, with coefficients which can be determined from the values of the elastic constants, the Peierls stresses, the properties of possible supplementary stable states, or other relevant experimentally determined mechanical quantities.

We mention explicitly that the potentials in (\ref{potz0}) describe lattices with isotropic elastic moduli. Anisotropic elasticities may be introduced (except for the hexagonal case, where they are forbidden by symmetry) by assuming a suitable separate dependence of the potentials on the real and imaginary parts of $J(z)-J(z_0)$ rather than on its modulus $J_0(z) = |J(z) - J(z_0)|$ as in (\ref{energyJ0})-(\ref{potz0}). For example, an energy exhibiting anisotropic square elasticities is obtained by writing:
\begin{equation}
\sigma_\text{sq,aniso}\big(\bCo\big)= \mu_1\; \text{Re}^2\sqrt{J(z)-1}+\mu_2 \;\text{Im}^2\sqrt{J(z)-1} \,,
\end{equation}
with $\mu_1,\mu_2$ positive moduli.

\section{Plastification under shearing}

We illustrate some relevant features of the model through the results of two numerical simulations of plastic-flow initiation in quasi-static conditions, for a square and a hexagonal lattice. To emphasize the basic effects of symmetry in these plastifying crystals, we consider the simplest case of loading each body along a primary-shear direction (i.e.~a densest crystallographic line) in the lattice.

The body is square-shaped, and contains an initially homogeneous material in which the underlying lattice is defect-free. The bottom side is aligned with a dense (primary-shear) lattice direction, in turn coinciding with one of the given orthonormal basis vectors. The strain-controlled shear boundary conditions are imposed to the top side with the bottom side kept fixed, while the remaining two sides are left stress-free. The boundary shear parameter is $\gamma \in \R$, with $\gamma =0$ in the ground state $z_0$ for both the square ($z_0  = i$) and the  hexagonal ($z_0=\rho$) case. For each $\gamma$ we search for the local minimizers of the body's total strain-energy functional computed through the density (\ref{eq:sigelastic})-(\ref{potz0}), and complying with the imposed boundary conditions.

The minimization algorithm follows a  preconditioned energy-gradient descent, implemented through a continuous FreeFEM code \citep{Hecht} using an unstructured mesh for a total of about $3\times 10^4$ degrees of freedom. At each step, the initial guess coincides with the metastable state associated with the previous value of the imposed loading. The algorithm then looks for metastable states which are close (in the sense of the preconditioned gradient-descent) to the previous equilibrium. The FEM discretization introduces in our non-convex problem a regularizing length scale related to lattice discreteness. Therefore, the mesh is fixed, and the solutions are the piecewise-differentiable functions with constant gradient over FEM cells and possible gradient jumps over mesh sides.

A well-known property of $j(z) \equiv 1728J(z)$ is that it has all integers in its Fourier-expansion coefficients \citep{modularforms2, modularforms1}:
\begin{equation}
j(\tau )=\frac{1}{q}+744+196884\,q+21493760\,q^{2}+864299970\,q^{3}+20245856256\,q^{4}+\dots,
\label{furierj}
\end{equation}
where $q = \exp(2\pi i\tau)$. In particular, we used the first 25 terms \citep{Jcoefficients} in (\ref{furierj}) for our minimization algorithm. The volumetric energy in (\ref{eq:sigelastic}) has been chosen of the form
\begin{equation}
\sigv(\det\bC) = \lambda (\det\bC-\log\det\bC),
\label{volumetric}
\end{equation}
with $\lambda/\mu\sim 30$ to enforce quasi-incompressibility.

\subsection{Shearing of a square lattice}

\begin{figure}

\centering

\includegraphics[width=\textwidth]{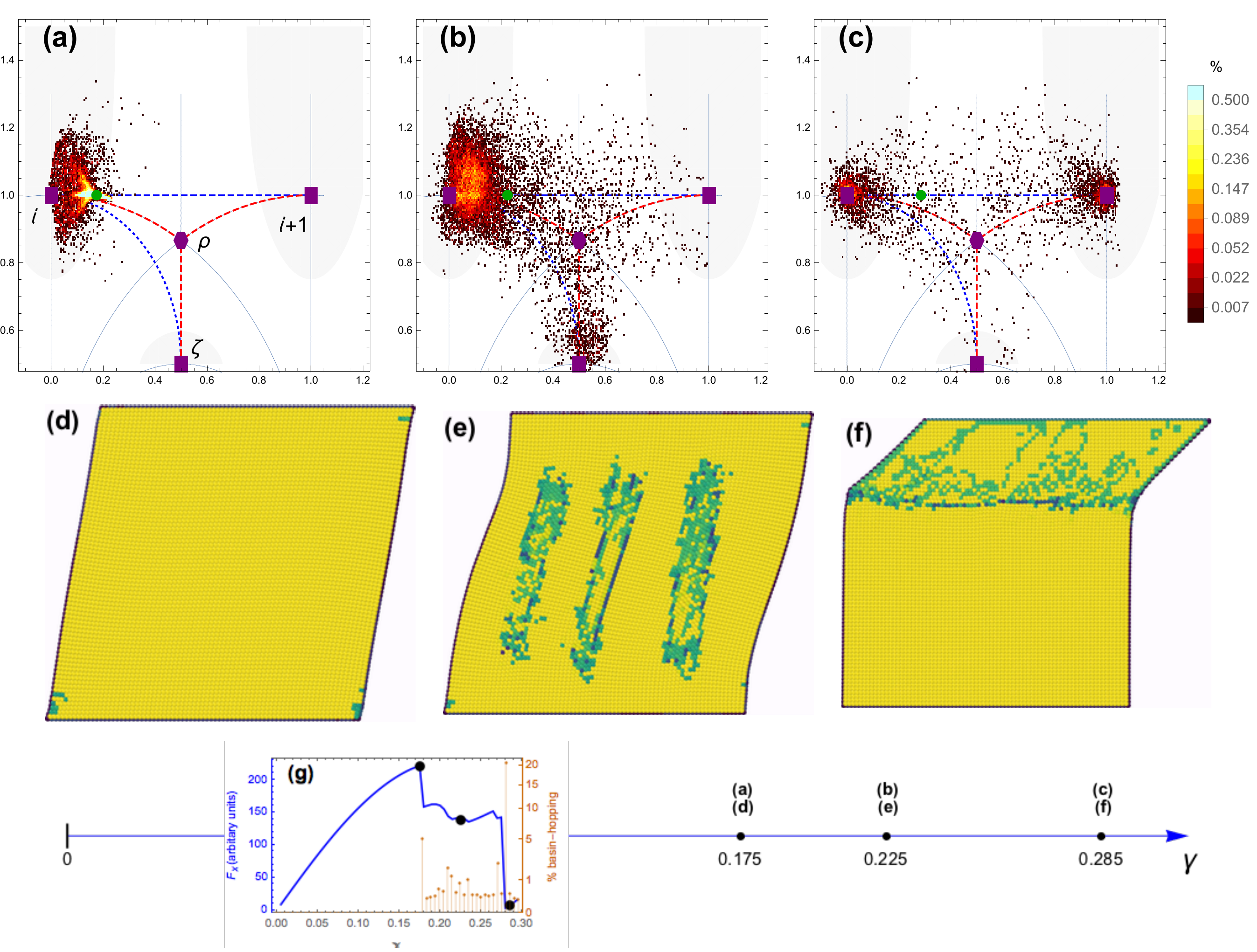}

\caption{(Color online) Simulation of plastic flow initiation in a square crystal on the Dedekind tessellation of $\mathbb{H}$, see Fig.~\ref{fig:Dedekind}. The imposed loading is along a primary shear direction in the lattice, parallel to the driven horizontal body-sides. The associated path in $\mathbb{H}$, parameterized by the increasing shear $\gamma$ (green dot) from $i$ to $i+1$, is the straight dashed blue line in panels (a),~(b),~(c). The optimal barrier-crossing path from $i$ to $i+1$ (i.e., the valley floor that will be studied in Section 6) is marked in dashed-red. It crosses the degenerate monkey saddle at the hexagonal point $\rho = e^{i \pi/3}$, where there is a bifurcation also to the optimal path from $i$ to $\zeta$ (also marked in dashed-red). Shading shows the convexity domains in $\mathbb{H}$ around the square minimizers of the energy (\ref{eq:ensq}). The three simulation snapshots (a),~(b),~(c) show the evolution of the strain clustering during plastification, given by the 2D-histogram for the $\bCo$-strain density on $\mathbb{H}$. Panels (d),~(e),~(f)  show the corresponding body-shape change. The colors highlight different centrosymmetry-parameter ranges, and defect evolution, in the lattice. Notice the defect pattern in the slip band in panel (f). Panel (g) displays the stress-strain relation (blue curve) during loading as a function of $\gamma$: the response is elastic to about $\gamma \simeq 0.17$, after which bursty plastic flow begins, as indicated by the intermittent percentage of strains that jump energy basin (orange spikes).
See the Supplementary Video V1 for more details.
}
\label{fig:snapshots}
\end{figure}

In the first simulation (see Fig.~\ref{fig:snapshots}) the energy $\hsigd$ in (\ref{eq:sigelastic}) is given by (\ref{potz0}) with $z_0=i$, $J(i) = 1$, and $\kappa(z_0)=2$, i.e.:
\begin{equation}
\sigma_{\text{d,sq}}=\mu \big|J(z)-1\big|.
\label{eq:ensq}
\end{equation}
The imposed primary-shear path $i \rightarrow 1+i$ is the straight horizontal dashed-blue line in Figs.~\ref{fig:snapshots}(a)-(b)-(c), with parameter $\gamma$ such that $\gamma =0$ in the ground state $z_0=i$, and $\gamma = 1$ in the neighboring equivalent sheared-square configuration $i+1$ (see Figs.~\ref{fig:Dedekind} and~\ref{fig:snapshots}(a)).

Figs.~\ref{fig:snapshots}(d)-(e)-(f) show three snapshots of the resulting $\gamma$-dependent plastification field. Figs.~\ref{fig:snapshots}(a)-(b)-(c) show the clustering of $\bCo$-values, which, via (\ref{bridge}) is represented by a $\gamma$-evolving cloud of points on the Dedekind tessellation of $\mathbb{H}$. This distribution of local-strain density gives a very informative picture of the main properties of the plastification field, see also \cite{PRBbarrera}, \cite{PRLgruppone}. The Supplementary Video V1 shows the entire simulation, reporting also the $\gamma$-dependent histograms for the local rotation angle $\theta$ and of $\det \bF$.
Figs.~\ref{fig:snapshots}(a)-(b)-(c) show the defects mediating plastic flow. They are heuristically identified through the centrosymmetry parameter used by the visualization software Ovito~\citep{ovito}, which counts the number of nearest neighbors of lattice points. In this setting dislocations emerge when nearby cells undergo concentrated slips originating from shears in GL$(2,\Z)$ which cannot be accommodated by elastic deformations. The stress distribution of a  trapped lattice defect in Figs.~\ref{fig:snapshots}(d)-(e)-(f) shows the correct features of dislocations' elastic field both far from and in the proximity of the core region, as in Fig.~4 in \cite{PRLgruppone}.

Fig.~\ref{fig:snapshots}(g) shows that the initially defect-free lattice goes through a significant elastic load-up with a quasi-homogeneous initial deformation. As $\gamma$ increases, the strain cloud in $\mathbb{H}$ widens due to the growing strain heterogeneity originating from the unloaded body-sides.

The end of the elastic regime is marked by a first large plastification event at about $\gamma \simeq 0.17$,
when the strain cloud suddenly jumps in $\mathbb{H}$ away from the initial point $i$, see Fig.~\ref{fig:snapshots}(b). This first large plasticity burst occurs as local strains cross over to \emph{both} the two symmetry-equivalent square energy-basins closest to $i$ in the direction of the imposed boundary condition, i.e.~$i +1$ and $\zeta$.

This complex plasticity mechanism originates from the way global GL-sym\-metry locally moulds the energy landscape, and how this interacts with the driving boundary condition. Both the square points~$i +1$ and $\zeta$ are involved because the optimal barrier-crossing path from $i$ to $i+1$  goes through the hexagonal point $\rho = e^{i \pi/3}$, which is a degenerate monkey saddle.\footnote{Since as noted earlier $J'(\rho) = J''(\rho) = 0$, for the square energy function $\sigma_{\text{d,sq}}$ in (\ref{eq:ensq}) there are degenerate monkey saddles in $\rho$ and in all other hexagonal points of $\mathbb{H}$, with null Hessian and third-order terms, up to a suitable rotation, of the form $x^3 - 3 xy^2$.}
There is thus in $\rho$ a bifurcation of the optimal square-to-square barrier-crossing path $i \rightarrow i+1$ also to the optimal path and $i\rightarrow \zeta$ going to the other nearby equivalent square point $\zeta$ (see the dashed-red fat-rhombic curves in Figs.~\ref{fig:snapshots}(a)-(b)-(c)).

Due to this bifurcation, plastic-flow initiation does not occur through the activation of the primary lattice shear $i \rightarrow i + 1$ (on the [0,1] lattice line) pertaining to the external driving $\gamma$ (dashed-blue line), but rather through a strain avalanche involving mainly the other primary shear path $i \rightarrow \zeta$ (dotted-blue line in Fig.~\ref{fig:snapshots}(a)), which is on the [1,0] line and is symmetry-related to $i \rightarrow i + 1$ (note that these two shear paths coincide in the linear approximation at $i$).
As a consequence, although the body is being loaded in the horizontal shear direction $i \rightarrow i +1$,  we observe the formation of the vertical slip bands in Fig.~\ref{fig:snapshots}(e) as a first complex event marking the onset of plastification in the square lattice (see also the Supplementary Video V1).
The width and shape of the associated strain cloud in Fig.~\ref{fig:snapshots}(b) actually indicates that not only are both the primary shears $i \rightarrow i+1$ and $i\rightarrow \zeta$ becoming simultaneously active as the imposed boundary condition follows the path $i \rightarrow i+1$, but that plastic flow indeed takes place with the concurrent activation of many other local deformation paths, with the elastic stabilization of strain values also belonging to the non-convexity region near the degenerate saddle in $\rho$ (see Fig.~\ref{fig:snapshots}(b)).\footnote{The locally-destabilizing shear directions originating from the loss of positive-definiteness of the acoustic tensor examined by \cite{PRLgruppone}  further contribute to specifying the bifurcation behavior of the energy-relaxing strain field. We notice there the creation of higher strain inhomogeneities and a finer spatial distribution of defects in the body already at the first nucleation (see Fig.~5 in \cite{PRLgruppone}), due to the imposition of hard boundary conditions. By contrast, in the present study we observe the creation of a broad-band defect pattern upon the first strain burst (Fig.~\ref{fig:snapshots}(a)-(b)). This is due the less stringent boundary constraints presently considered, involving free body sides.}

A large stress drop corresponds to this first relaxation event, see Fig.~\ref{fig:snapshots}(g), giving rise to the nucleation peak in the stress-strain relation, as expected in this originally clean system \citep{cottrell49fedelich92truvain041, cottrell49fedelich92truvain042, cottrell49fedelich92truvain043}. The orange spikes in Fig.~\ref{fig:snapshots}(g) show that after the first event a discontinuous strain activity is ever present in the body due to the coordinated basin hopping of local strain values. Correspondingly, intermittent defect nucleation and dynamics spontaneously progress in the lattice by energy minimization, with no need for extra assumptions, see also the Supplementary Video V1. Bursty plastic flow ensues, as is typically observed in the microscale plasticity of crystalline materials \citep{zapperi02, dimiduk06, zapperisciencebursts, uchic092, papanikolau}.

Under a further shear increase the body eventually separates (for $\gamma \simeq 0.28$) with a large plastic slip into two clearly defined regions in Fig.~\ref{fig:snapshots}(f), one almost undistorted and the other deformed in the direction of the boundary shear, at the higher shear value pertaining to the minimizer $i + 1$. The width of the slip band is dictated by the shear value $\gamma$ imposed at the boundary. This large slip event is again associated to a marked stress relaxation, see the second large drop in the stress-strain curve in Fig.~\ref{fig:snapshots}(g).

Of particular interest in Fig.~\ref{fig:snapshots}(f) is the defect pattern produced within the slip band. This band originates as points in the upper part of the body shear to $i+1$ from both near $i$ and near $\zeta$, rather than directly from $i$, see Figs.~\ref{fig:snapshots}(b)-(c) and the Supplementary Video V1. When neighboring lattice cells reach the same strain-energy well $i+1$ in $\mathbb{H}$ through the paths
\begin{equation}
i \rightarrow \zeta \rightarrow i + 1 \qquad  \text{vs.} \qquad  i \rightarrow i + 1,
\label{differentpaths}
\end{equation}
different values in their local rotation angle $\theta$ may occur (and are energetically cost-free due to Galilean invariance~(\ref{strainenergy})). Such rotation differences in nearby cells experiencing the same strain $i+1$ create kinematic incompatibilities which give rise to the slip-band defects shown in Fig.~\ref{fig:snapshots}(f), see also \cite{biurzaza}, \cite{gao1}.

We have clear evidence here of the importance of the bifurcations on the paths followed by nearby local strains while changing basin driven on the GL-topography. These bifurcations behave as `disorder engines', enhancing defect creation in the body, in the first nucleation events and all along the flow. The basic energy $\sigma_{\text{d,sq}}$ in (\ref{eq:ensq}) presents such effects in an archetypical way for square lattices, due to the bifurcations occurring through the degenerate monkey saddles at the hexagonal configurations.

\subsection{Shearing of a hexagonal lattice}

\begin{figure}

\centering
\includegraphics[width=\textwidth]{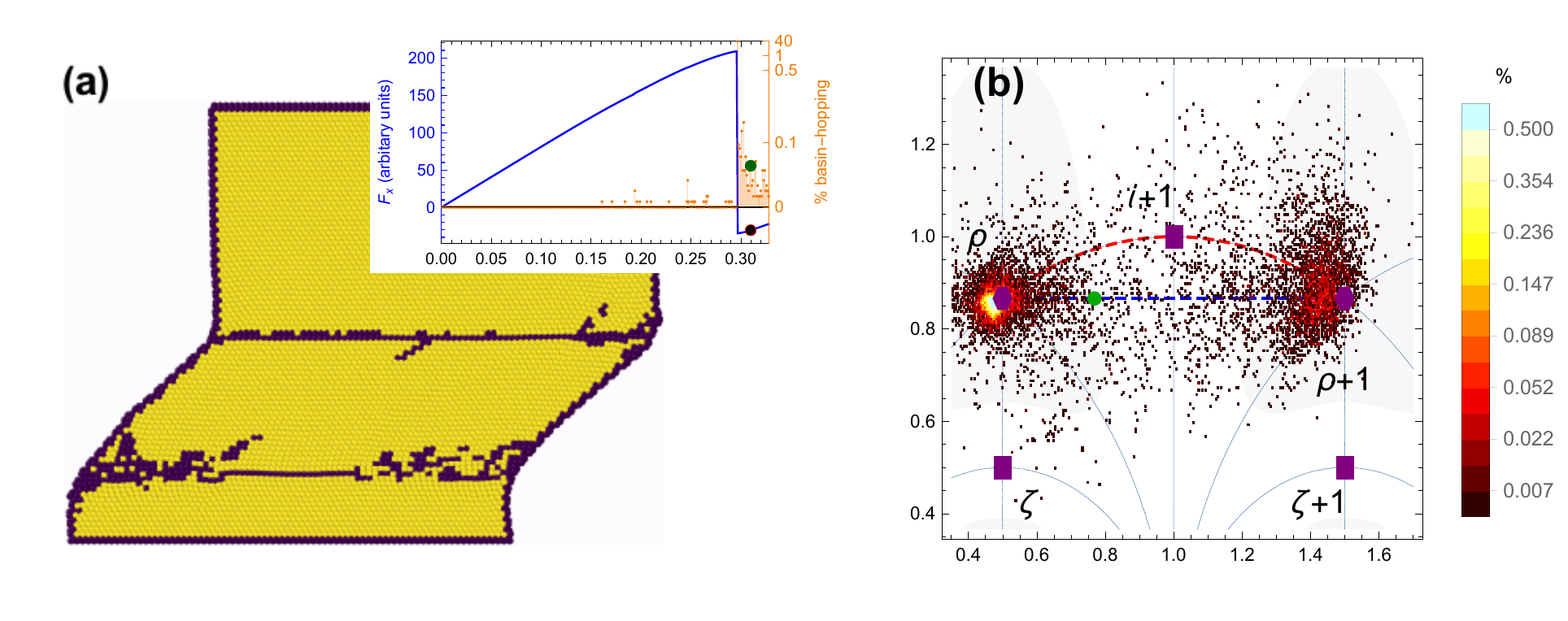}

\caption{(Color online) Simulation of plastic flow initiation in a hexagonal crystal.
The imposed primary-shear path is from $\rho$ to $\rho +1$ (see Fig.~1), with same details as in Fig.~2. (a)~Snapshot of the body-shape and lattice-defect configuration after the first large plastification event at $\gamma \simeq 0.29$.
The inset shows the stress-strain relation (blue curve) and bursty plastic flow activity (in orange) upon loading. See the Supplementary Video V2 for more details. In contrast to the square case, notice here the formation of a largely defect-free slip band in the body.  (b) The corresponding strain clustering on $\mathbb{H}$. The optimal barrier crossing path is red-dashed through the square configuration, while the dashed blu line indicates the primary shear path.}
\label{fig:snapshothexagonal}
\end{figure}

The second simulation is performed, under the same conditions as the previous one, for an initially defect-free hexagonal crystal with ground state $\rho$ (Fig.~1). The energy $\hsigd$ in (\ref{eq:sigelastic}) is now given by (\ref{potz0}) for $z_0=\rho$, $J(\rho) = 0$, and $\kappa(z_0)=3$:
\begin{equation}
\sigma_{\text{d,hex}}=\mu \big|J(z)\big|^{2/3}.
\label{eq:enhex}
\end{equation}
As in the square case, we see that after the initial elastic load-up, plastic flow initiates through a large plastification avalanche.  Figs.~\ref{fig:snapshothexagonal}(a)-(b) present a snapshot of the corresponding body shape and strain cloud in $\mathbb{H}$.  The inset in Fig.~\ref{fig:snapshothexagonal}(a) shows the strong stress relaxation associated to this first strain event, and the bursty character of the ensuing plastic flow, as indicated by the orange strain-activity spikes. Full results are reported in the Supplementary Video V2.

Unlike with the square case discussed above, plasticity in this hexagonal lattice largely follows the primary-shear path imposed by the boundary conditions on the body. This is again a direct consequence of the GL-symmetry constraints. Indeed, for the hexagonal energy $\sigma_{\text{d,hex}}$  in (\ref{eq:enhex}), with minimizers on the hexagonal points $\rho$, $\rho+1$, $...$, the saddles at the square points $i$, $i+1$, $\zeta$, $...$, are necessarily standard, with an indefinite non-degenerate Hessian. For instance, in Fig. 3(b) the square saddle in $i+1$ is the mountain pass for the red-dashed hexagonal-to-hexagonal optimal path $\rho \rightarrow \rho+1$ (the dashed blu line indicates the primary shear path). In general, each square saddle is here the mountain pass on a unique hexagonal-to-hexagonal optimal path, with no barrier-crossing bifurcations. We remark that in this case the slipped portion of the crystal is largely defect free, see Fig.~\ref{fig:snapshothexagonal}(a), unlike with the square case in Fig.~\ref{fig:snapshots}(f).

\section{Networks of energy extremals and valley floors: Bethe trees and Husimi cactuses}

The simulations in the previous Section extend the observations in \cite{PRLgruppone} concerning the complexity of collective defect nucleation in the present plasticity framework. They especially highlight the role played in crystal plasticity by the symmetry-imposed constraints on the energy extremals and on the basin-hopping strain paths on $\mathbb{H}$, with their possible bifurcations. In particular, we see such paths are largely directed along the valley floors of the energy landscape. To better elucidate these effects, we explicitly consider the gradient-extremal curves on $\mathbb{H}$ pertaining to our energy functions. These are loci defined by the condition (\ref{valleyfloors})$_1$ that $\bna \sigma$ be an eigenvector of $\bH$ \citep{valleyfloor1, valleyfloor2, valleyfloor3}, where $\bna\sigma$  and $\bH$ are respectively the gradient and the Hessian on $\mathbb{H}$ of the energy $\sigma$. Inequality (\ref{valleyfloors})$_2$  then identifies the valley floors of the energy topography, as the gradient-extremal curves occupying the energy landscape bottom:
\begin{equation}
\bH\,\bna\sigma  - \xi\, \bna\sigma  =0, \; \; \xi \in \R, \quad \text{and} \quad \bH\be\cdot\be\geq 0,
\label{valleyfloors}
\end{equation}
where $\be$ is a vector orthogonal to the energy gradient in the hyperbolic metric.
The arrangement of these energy-surface features is summarized by an infinite GL-invariant network whose nodes are the energy extremals and edges are along the energy valley floors of $\sigma$ (see also related considerations by \cite{denoualJMPLplasticity}, \cite{gao1}). The node-coordination values $c$  on this network, with the corresponding edge multiplicity, are determined by GL-symmetry. In particular, nodes with $c>2$ give valley-floor bifurcations. The topological and metric features of this network underpin a crystal's plastification mechanisms. They make explicit the geometric scaffolding on which the GL-energetics largely leads, via the short and long range elastic interactions, the coordinated local strain-jump activity on $\mathbb{H}$ producing the plastification bursts, with the boundary conditions contributing to select the activated deformation paths.

\begin{figure}

\centering

\includegraphics[width=\textwidth]{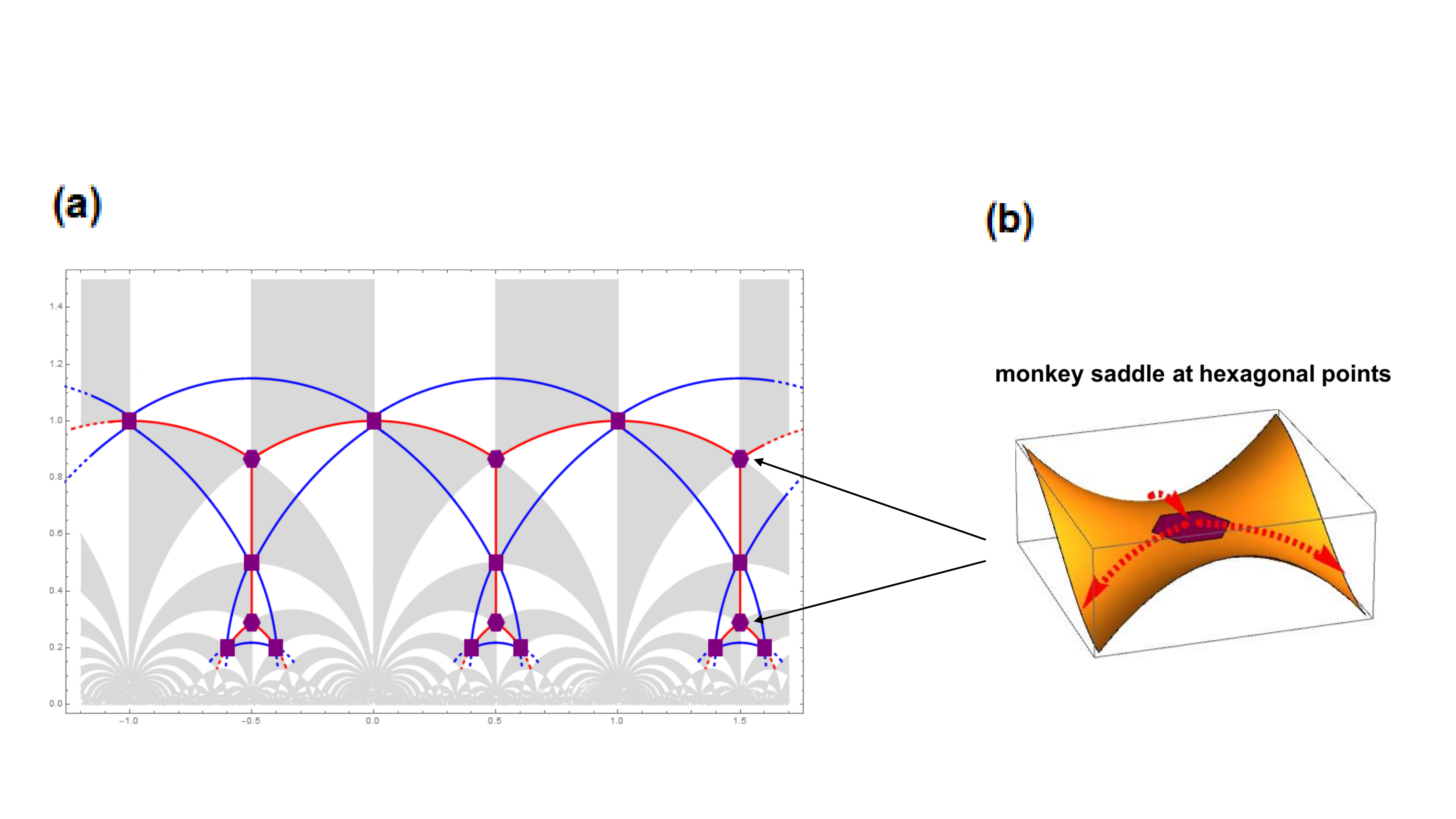}

\caption{(Color online)  The space $\mathbb{H}$ with highlighted in red a portion of the infinite graph whose nodes are on the square and hexagonal points, and edges on fat-rhombi curves. (a) Due to GL-symmetry, this network is a Bethe-like tree, with coordination $c=3$ in $\rho$ and all the other hexagonal nodes, and decorated by further nodes $i$, $\zeta$, ..., with coordination $c=2$, located at the edges' mid-points. On this geometric scaffolding are located the extremals and valley floors for both the strain-energies $\sigma_{\text{d,sq}}$ and $\sigma_{\text{d,hex}}$ in (\ref{eq:ensq})-(\ref{eq:enhex}). Indicated in blue is also the network containing isolated loops (a Husimi cactus) as discussed in the text.
(b) In the case of the square energy $\sigma_{\text{d,sq}}$ in (\ref{eq:ensq}), the associated network coincides with the Bethe tree in (a), whereon the 3-coordinated hexagonal points are all monkey saddles bifurcating the valley floors of the GL-landscape, and the 2-coordinated well bottoms are at the square points located mid-way between two adjacent saddles. The position of two monkey saddles (b) is indicated by the arrows.
}
\label{fpaths}
\end{figure}

In particular, for both the square and hexagonal energies  (\ref{eq:ensq}) and  (\ref{eq:enhex}) considered above, the valley-floor network coincides with the Bethe-like tree \citep{husimi1, husimi2} shown in red in Fig.~\ref{fpaths}(a).\footnote{The fact that the valley-floor networks for both our energies(\ref{eq:ensq}) and  (\ref{eq:enhex}) coincide with the Bethe tree in Fig.~\ref{fpaths}(a) whose edges are on the boundary of $\cal D$, can be derived directly from the GL-symmetry relations across each boundary portion of $\cal D$ and the monotonic growth properties of  $J$ on $\partial\cal D$. A more familiar representation of this Bethe tree is obtained when considering a disc model of the hyperbolic plane, as in \cite{ContiZanzotto} or \cite{tesipatriarca}. See also \cite{bethesymmetree}.}
This is characterized by 3-connected hexagonal nodes, with further 2-connected square nodes mid-way along the edges, given by the fat-rhombic valley floors of both (\ref{eq:ensq}) and (\ref{eq:enhex}). The plastification and defect-generation scenarios for square vs.~hexagonal lattices in Section~5 originate from the different positions of the saddles on this Bethe tree. In the hexagonal case (\ref{eq:enhex}) the (standard) saddles are located at non-bifurcating nodes with $c=2$. Valley-floor bifurcations only take place here at the 3-connected minimizer (hexagonal) nodes, with local plastic strains largely confined near the primary-slip paths (see Fig. 4(b)). By contrast, for the square energy (\ref{eq:ensq}) the (degenerate) saddles coincide with the 3-connected bifurcating nodes of the Bethe tree. This promotes the complex deformation mechanisms involving the conjunct activation of several plastification paths at once and sequentially in the body.

It is interesting to briefly examine also the structure of the valley-floor networks  associated to more generic GL-invariant strain potentials (\ref{eq:energyinvariance})-(\ref{energyJ}) than the simplest ones (\ref{eq:ensq})-(\ref{eq:enhex}) considered above. A family of twice-differentiable GL-energies produced by \cite{ContiZanzotto} allows for a one-parameter unfolding of the hexagonal monkey saddles of $\sigma_{\text{d,sq}}$ in  (\ref{eq:ensq}), see the square-hexagonal minimizers' bifurcation diagram in Fig.~4 of that work. GL-invariant  $J$-energies with an analogous square-hexagonal bifurcation are also considered in \cite{tesipatriarca}, as a linear combinations of $\sigma_{\text{d,sq}}$ and $\sigma_{\text{d,hex}}$ in (\ref{eq:ensq}) and (\ref{eq:enhex}). As discussed in \cite{PZbook}, the bifurcaton features, with the related unfolding of the possibly degenerate extremals, depend on the symmetry properties of the bifurcation points: in the present case for $i$ (square) we have a subcritical pitchfork to fat-rhombic configurations, and for $\rho$ (hexagonal) the bifurcation is transverse to three rhombic branches, see \cite{ContiZanzotto}.

As a consequence, the hexagonal monkey saddles first unfold generically into 'monkey regions' involving three standard saddles at fat [skinny] rhombic points in the vicinity of a hexagonal minimum [maximum] of the energy,\footnote{For a two-parameter unfolding of the monkey saddle and the corresponding phase diagram, see for instance \cite{electronsmonkeysaddle}.} see also Fig.~6 in \cite{PRLgruppone}. Plastification mechanisms and defect generation are largely unaffected by these small deviations from the degenerate monkey-saddle configuration.

On the other hand, when the unfolded extremals are skinny-rhombic saddles located away from the hexagonal maximizer,  these become 2-connected (standard) mountain passes each between a suitable pair of minimizing square nodes. This is analogous to what happens in the hexagonal case with $\sigma_{\text{d,hex}}$ in (\ref{eq:ensq}), where only 2-connected standard-saddle nodes are present, and no valley-floor bifurcations other than at the minimizers. However, the valley-floor network topology associated to this new configuration of unfolded extremals is very different from the previous cases, as it now involves triangle-loops with vertices on the 4-connected square minimizers. The resulting graph is thus a Husimi cactus \citep{husimi1, husimi2}, rather than a loop-free Bethe tree, see the blue network in Fig.~\ref{fpaths}(a). This points to two distinct scenarios for plastification and defect generation in square lattices. By contrast, the bifurcation diagram shows there is a single scenario possible for GL-energies with hexagonal ground states, as already described above.

We conclude that interesting insight on the behavior of crystalline materials is gleaned from the energy-surface features summarized by the associated network of extremals and valley floors. Analogous studies of energy features are done in other fields, as in organic chemistry and mechanochemistry \citep{saddleinflections3, e2005, e2007, saddleinflections2, saddleinflections4, saddleinflections5}, protein folding and other biomolecules' dynamics \citep{monkeybiomolecules, proteinfolding}, optics \citep{optics}, or superconductivity \citep{electronsmonkeysaddle}. The techniques for energy-topography analysis there developed will thus help shed light also on crystal mechanics phenomena (see also the graph analysis methods in \cite{morse}).
In particular, taken together with the results by \cite{PRLgruppone}, the discussion above concerning different plasticity scenarios for different lattice symmetries should aid in the understanding of the complex behavior and marked dissimilarities experimentally observed in the microplasticity of materials with distinct crystal structures, including sensitivity to orientation and loading type, as well as defect-pattern formation and non-universal scaling exponents for dislocation avalanches, as referenced in the Introduction.

\newpage

\section*{APPENDIX}

\section*{Caption to Supplementary Video V1}

Numerical simulation of the plastic flow initiation in a homogeneous square body containing an initially defect-free square lattice. The imposed loading is along a primary shear lattice direction, aligned with the parallel square-cell side and bottom body side. The shearing is imposed through the constrained horizontal sides of the body, with the two remaining sides free. The strain-energy is given by Eq.~(\ref{eq:ensq}). See also Figs.~1-2 in the main text.

{\bf (a)} Bursty deformation field in the body for increasing shear parameter $\gamma$, indicated by the moving green dot along the $\gamma$-axis in {\bf (e)}. Lattice points are color-coded according to {\bf (c)} depending on the energy basin in the Poincar\'e half-plane $\mathbb{H}$ visited by the strain of the associated lattice cell during loading. Notice the defect dynamics intrinsically produced in this model by energy minimization.

{\bf (b)}-{\bf (d)}-{\bf (f)} Intermittent evolution of the $\gamma$-dependent histograms of the deformation-gradient parameters recorded during the simulation.

{\bf (b)} Evolution of the histogram (strain cloud) of the density of strain parameters on the Dedekind tessellation of $\mathbb{H}$ during loading. The histogram color-coding provides the percentage of body FEM cells with strain at each point of $\mathbb{H}$. The straight horizontal dashed-blue line between the two neighboring square configurations $i$ and $i+1$ is the image in $\mathbb{H}$ of the primary shear path imposed as boundary condition. The initial configuration is in $i$ for $\gamma = 0$, while $\gamma = 1 $ corresponds to $i+1$. The energetically optimal barrier-crossing path from $i$ to $i+1$ is the red-dashed path passing through the hexagonal monkey saddle in $\rho= e^{i \pi /3}$. Such path bifurcates in $\rho$ also to the optimal barrier-crossing path from $i$ to $\zeta= \frac{1}{2}(i+1)$ (the blue-dotted line indicates the corresponding primary shear path from $i$ to $\zeta$). Shading shows the convexity domains around the hexagonal minimizers in $\mathbb{H}$ of the square energy function.

{\bf (c)} Color-code map used in {\bf (a)} for the (symmetry-related copies of the fundamental domain $\cal D$ which make up the various) energy-basins on $\mathbb{H}$ pertaining to the square strain-energy in Eq.~(17) of the main text.

{\bf (d)} Evolution of the histogram for the values of $\det \bF$, indicating volumetric effects in the lattice.

{\bf (e)} Bursty plastic flow. Jagged stress-strain behavior of the body (blue), and the associated spikes (orange) showing the percentage of basin-hopping strain values during loading, as $\gamma$ grows. The body response is elastic to about  $\gamma \simeq 0.17$, after which bursty plastic flow begins, with a large relaxation event, continuing then for growing $\gamma$. A second substantial plastification takes place at $\gamma \simeq 0.28$, where the body separates, in a large plastic slip, into two clearly defined regions.

{\bf (f)} Evolution of the histogram for the values of the angle $\theta$ in the polar decomposition of the deformation gradient $\bF$, indicating local rotation in the lattice.

Notice in \textbf{(d)} and \textbf{(f)} the effects associated with the slip band formation. Both the local dilations and local rotations relax to almost homogeneous values after the second plastification event.  In particular, the rotation-angle $\theta$ distribution is bimodal, with peaks corresponding to the (zero) rotation in the undistorted portion of the body and to the value of $\theta$ pertaining to the complete shear induced by the boundary condition, concentrated in the slip band.

\section*{Caption to Supplementary Video V2}

Numerical simulation of plastic flow initiation in a homogeneous square body containing an initially defect-free hexagonal lattice. The imposed loading is along a primary shear lattice direction, aligned with a hexagonal-cell side parallel to the bottom body side.
The shearing is imposed through the constrained horizontal sides of the body, with the two remaining sides free. The strain-energy is given by Eq.~(\ref{eq:enhex}). See also Figs.~1 and 3 in the main text. All panels for this Supplementary Video V2 refer to the same variables and distributions as in the caption to Supplementary Video V1.

In this case the slip-band formation results from a single large plastification avalanche for $\gamma\simeq 0.29$, close to the value where we observe the slip-band formation also in the square lattice in Supplementary Video V1. The cells involved in the present slip event, driven by the boundary conditions, jump directly from the initial well at $z=\rho$ to the neighboring well at $z=\rho+1$, without the bifurcation activating other paths as observed in the square case of Supplementary Video V1. This limits the number of defects created in the slip process, and the resulting slip-band is largely defect-free, see also Fig.~3 of the main text.


\bigskip

\begin{thebibliography}{99}

\bibitem[Alava et al., 2014]{zapperiDDDreview}
Alava, M.J., Laurson, L., Zapperi, S., 2014. Crackling noise in plasticity. Eur.\ Phys.\ J.\ Sp.\ Top.\ 223:2353-2367.

\bibitem[Alkan et al., 2018]{nonschmid2}
Alkan, S., Ojha, A., Sehitoglu, H., 2018. The complexity of non-Schmid behavior in the CuZnAl shape memory alloy. J.\ Mech.\ Phys.\ Solids 114:238-257.

\bibitem[Apostol, 1976]{modularforms1}
Apostol, T.M., 1976. Modular functions and Dirichlet series in number theory, Springer-Verlag, Berlin.

\bibitem[Arriaga and Waisman, 2017]{arriaga17}
Arriaga, M., Waisman, H., 2017. Combined stability analysis of phase-field dynamic fracture and shear band localization. Int.\ J.\ Plast. 96:81-119.

\bibitem[Baggio et al., 2019]{PRLgruppone}
Baggio, R., Arbib, E., Biscari, P., Conti, S., Truskinovsky, L., Zanzotto, G., Salman, O.U., 2019. Landau-type theory of planar crystal plasticity, arXiv:1904.03429 [cond-mat.mtrl-sci].

\bibitem[Bakst et al., 2018]{bakst18}
Bakst, I.N., Yu, H., Bahadori, M., Yu, H., Lee, S.-W., Aindow, M., Weinberger, C.R., 2018. Insights into the plasticity of Ag3Sn from density functional theory. Int.\ J.\ Plast. 110:57-73.

\bibitem[Balandraud et al., 2015]{PRBbarrera}
Balandraud, X., Barrera, N., Biscari, P., Gr\'ediac, M., Zanzotto, G., 2015. Strain intermittency in shape-memory alloys. Phys.\ Rev.\ B 91:174111.

\bibitem[Bhattacharya, 2004]{bhattabook}
Bhattacharya, K., 2004. Microstructure of Martensite: Why It Forms and How It Gives Rise to the Shape-Memory Effect, Oxford University Press.

\bibitem[Bhattacharya et al., 2004]{BCZZnature}
Bhattacharya, K., Conti, S., Zanzotto, G., Zimmer, J., 2004. Crystal Symmetry and the reversibility of martensitic transformations. Nature 428:55-59.

\bibitem[Bhattacharya and James, 2005]{jamesperspective1}
Bhattacharya, K., James, R.D., 2005. The Material Is the Machine. Science 307:53.

\bibitem[Biscari et al., 2015]{biurzaza}
Biscari, P., Urbano, M.F., Zanzottera, A., Zanzotto, G., 2015. Intermittency in crystal plasticity informed by lattice symmetry. J. Elasticity 123:85-96.

\bibitem[Bociort and van Turnhout, 2005]{optics}
Bociort, F., van Turnhout, M., 2005. Generating saddle points in the merit function landscape of optical systems, Proc.\ SPIE 5962:0S1-8.

\bibitem[Brinckmann et al., 2008]{greer}
Brinckmann, S., Kim, J.Y., Greer, J.R., 2008. Fundamental differences in mechanical behavior between two types of crystals at the nanoscale. Phys.\ Rev.\ Lett. 100:155502.

\bibitem[Caspersen et al., 2004]{ironshear1}
Caspersen, K., Lew, A., Ortiz, M., Carter, A., 2004. Importance of shear in the bcc-to-hcp transformation in iron. Phys.\ Rev.\ Lett. 93:115501.

\bibitem[Chan et al., 2010]{phasefieldcrystaldahmen}
Chan, P.Y., Tsekenis, G., Dantzig, J.,  Dahmen, K.A., Goldenfeld, N., 2010. Plasticity and dislocation dynamics in a phase field crystal model. Phys.\ Rev.\ Lett. 105:015502.

\bibitem[Cho et al., 2018]{nonschmid1}
Cho, H., Bronkhorst, C.A., Mourad, H.M., Mayeur, J.R., Luscher, D.J., 2018. Anomalous plasticity of body-centered-cubic crystals with non-Schmid effect. Int.\ J.\ Solids Struct. 139-140:138-149.

\bibitem[Chung and Lee, 2018]{bookplasticitylee}
Chung, K., Lee, M.-G., 2018. Basics of Continuum Plasticity. Ed.\ Springer, Singapore.

\bibitem[Conti and Zanzotto, 2004]{ContiZanzotto}
Conti, S., Zanzotto, G., 2004. A Variational Model for Reconstructive Phase Transformations in Crystals, and their Relation to Dislocations and Plasticity. Arch.\ Rational Mech.\ Anal. 173:69-88.

\bibitem[Csikor et al., 2007]{zapperisciencebursts}
Csikor, F.F., Motz, C., Weygand, D., Zaiser, M., Zapperi, S., 2007. Dislocation Avalanches, Strain Bursts, and the Problem of Plastic Forming at the Micrometer Scale. Science 318:251-254.

\bibitem[Cui et al., 2017]{differenceloading}
Cui, Y., Po, G., Ghoniem, N.M., 2017. Influence of loading control on strain bursts and dislocation avalanches at the nanometer and micrometer scale. Phys.\ Rev.\ B 95:064103.

\bibitem[de Miranda-Neto and Moraes, 1992]{husimi1}
de Miranda-Neto, J., Moraes, F., 1992. Symmetry properties of the Bethe lattice and the Husimi cactus. J.\ Physique I 2:1657-1666.

\bibitem[de Miranda-Neto and Moraes, 1993]{husimi2}
de Miranda-Neto, J., Moraes, F., 1993. Metric properties of the Bethe lattice and the Husimi cactus. J.\ Physique I 3:29-42.

\bibitem[Denoual et al., 2010]{corridoifrancesi}
Denoual, C., Caucci, A.M., Soulard, L., Pellegrini, Y.-P., 2010. Phase-field reaction-pathway kinetics of  martensitic transformations in a model Fe$_3$Ni alloy. Phys.\ Rev.\ Lett. 105:035703.

\bibitem[Dimiduk et al., 2006]{dimiduk06}
Dimiduk, D.M., Woodward, C., LeSar, R., Uchic, M.D., 2006. Scale-Free Intermittent Flow in Crystal Plasticity. Science 312:1188-1190.

\bibitem[Ding et al., 2012]{cottrell49fedelich92truvain043}
Ding X., Zhao, Z., Lookman, T., Saxena, A., Salje, EK., 2012. High junction and twin boundary densities in driven dynamical systems. Adv.\ Mater. 24:5385-5389.

\bibitem[Dmitriev et al., 1988]{transcendental1}
Dmitriev, V.P., Rocha, S.B., Gufan, Yu.M., Toledano, P., 1988. Definition of a transcendental order parameter for reconstructive phase transitions. Phys.\ Rev.\ Lett. 60:1958-1961.

\bibitem[E et al., 2005]{e2005}
E, W., Ren, W., Vanden-Eijnden, E., 2005. Transition pathways in complex systems: Reaction coordinates, isocommittor surfaces, and transition tubes. Chem.\ Phys.\ Lett. 413:242-247.

\bibitem[E et al., 2007]{e2007}
E, W., Ren, W., Vanden-Eijnden, E., 2007. Simplified and improved string method for computing the minimum energy paths in barrier-crossing events. J.\ Chem.\ Phys. 126:164103.

\bibitem[Edelsbrunner, 2001]{morse}
Edelsbrunner, H., Harer, J., Zomorodian, A., 2001. Hierarchical Morse-Smale Complexes for Piecewise Linear 2-Manifolds. Proc.\ 17th Symp.\ Comp.\ Geom. 70-79.

\bibitem[Ericksen, 1977]{eri77}
Ericksen, J.L., 1977. Special topics in elastostatics. Adv.\ Appl.\ Mech. 17:189-244.

\bibitem[Ericksen, 1980]{eri80}
Ericksen, J.L., 1980. Some phase transitions in crystals. Arch. Rational Mech. Anal. 73:99-124.

\bibitem[Ess et al., 2008]{saddleinflections2}
Ess, D. H., Wheeler, S.E, Iafe, R.G, Xu, L., Celebi-Olç{\"u}m, N., Houk, K.N., 2008. Bifurcations on potential energy surfaces of organic reactions. Angew.\ Chem.\ Int.\ Ed.  47:7592-7601.

\bibitem[Fedelich and Zanzotto, 1992]{cottrell49fedelich92truvain041}
Fedelich, B., Zanzotto, G., 1992. Hysteresis in discrete systems of possibly interacting elements with a double-well energy. J.\ Nonlin.\ Sci.\ 2:319.

\bibitem[Feng et al., 2019]{james2D}
Feng, F., Plucinsky, P., James, R.D., 2019. Phase transformations and compatibility in helical structures slip and twin, arXiv:1809.06282 [cond-mat.soft].

\bibitem[Folkins, 1991]{folkins}
Folkins, I., 1991. Functions of two-dimensional Bravais lattices. J.\ Math.\ Phys. 32, 1965-1969.

\bibitem[Freddi and Royer-Carfagni, 2016]{freddi16}
Freddi, F., Royer-Carfagni, G., 2016. Phase-field slip-line theory of plasticity. J.\ Mech.\ Phys.\ Solids 94:257-272.

\bibitem[Fressengas et al., 2009]{fressengeas09}
Fressengeas, C., Beaudoin, A.J., Entemeyer, D., Lebedkina, T., Lebyodkin, M., Taupin, V., 2009. Dislocation transport and intermittency in the plasticity of crystalline solids. Phys.\ Rev.\ B 79:014108.

\bibitem[Gao et al., 2019]{gao1}
Gao, Y., Wang, Y., Zhang, Y., 2019. Deformation pathway and defect generation in crystals: a combined group theory and graph theory description. Int.\ Union Cryst.\ J. 6:96-104.

\bibitem[Gurtin et al., 2010]{gurtin}
Gurtin, M.E., Fried, E., Anand, L., 2010. The Mechanics and Thermodynamics of Continua. Cambridge Univ.\ Press, New York.

\bibitem[Harlow et al., 2011]{bethesymmetree}
Harlow, D., Shenker, S., Stanford, D., Susskind, L., 2011. Eternal Symmetree, arXiv:1110.0496 [hep-th].

\bibitem[Healy and Ackland, 2014]{acklandasymmetry}
Healy C.J., Ackland, G.J., 2014. Molecular dynamics simulations of compression–tension asymmetry in plasticity of Fe nanopillars. Acta Materialia 70:105-112.

\bibitem[Hecht, 2012]{Hecht}
Hecht, F., 2012. New development in FreeFem++. J.\ Num.\ Math. 20:251-266.

\bibitem[Hoffman et al., 1986]{valleyfloor1}
Hoffman, D.K., Nord, R.S., Ruedenberg, K., 1986. Gradient Extremals. Theor.\ Chim.\ Acta 69:265-279.

\bibitem[Horovitz et al., 1989]{transcendental2}
Horovitz, B., Gooding, R.J., Krumhansl, J.A., 1989. Order parameters for reconstructive phase transitions. Phys.\ Rev.\ Lett. 62:843.

\bibitem[Irastorza-Landa et al., 2016]{dislocationstructures2}
Irastorza-Landa, A., Van Swygenhoven, H., Van Petegem, S., Grilli, N., Bollhalder, A., Brandstetter, S., Grolimund, D., 2016. Following dislocation patterning during fatigue. Acta Mater. 112:184-193.

\bibitem[James, 2015]{jamesperspective2}
James, R.D., 2015. Taming the temperamental metal transformation. Science 348:968.

\bibitem[James, 2018]{jamesreview}
James, R.D., 2018. Materials from mathematics. Bull.\ Amer.\ Math.\ Soc.\ 56:1-28.

\bibitem[Kamimura et al., 2018]{kamimura18}
Kamimura, Y., Edagawa, K., Iskandarov, A.M., Osawa, M., Umeno, Y., Takeuchi, S., 2018. Peierls stresses estimated via the Peierls-Nabarro model using \emph{ab-initio} $\gamma$-surface and their comparison with experiments. Acta Mater. 148:355-362.

\bibitem[Kilford, 2008]{poincarehalfplane2}
Kilford, L., 2008. Modular Forms - A Classical and Computational Introduction, Imperial College Press.

\bibitem[Kirsten and Williams, 2010]{modularphysics}
Kirsten, K., Williams,  F. L., 2010. A Window Into Zeta and Modular Physics, MSRI Publications, Vol. 57.

\bibitem[Kocks and Mecking, 2003]{dislocationstructures1}
Kocks, U.F., Mecking, H., 2003. Physics and phenomenology of strain hardening: the FCC case. Prog.\ Mater.\ Sci. 48:171-273.

\bibitem[Levitas, 2013]{levitas1}
Levitas, V.I., 2013. Phase-field theory for martensitic phase transformations at large strains. Int.\ J.\ Plast. 49:85-118.

\bibitem[Levitas, 2018]{levitas2}
Levitas, V.I., 2018. Phase field approach for stress- and temperature-induced phase transformations that satisfies lattice instability conditions. Part I. General theory. Int.\ J.\ Plast. 106:164-185.

\bibitem[Lew at al, 2006]{ironshear2}
Lew, A., Caspersen, K., Carter, E.A., Ortiz, M., 2006. Quantum mechanics based multiscale modeling of stress-induced phase transformations in iron, J.\ Mech.\ Phys.\ Solids 54:1276-1303.

\bibitem[Mallamace et al., 2016]{proteinfolding}
Mallamace, F., Corsaro, C., Mallamace, D., Vasi, S., Vasi, C., Baglioni, P., Buldyrev, S. V., Chen, S. H., Stanley, H. E., 2016. Energy landscape in protein folding and unfolding of proteins. Proc.\ Nat.\ Acad.\ Sci. 113:3159-3163.

\bibitem[McFaul et al., 2019]{dahmen19}
McFaul, L.W., Wright, W.J., Sickle, J., Dahmen, K.A., 2019. Force oscillations distort avalanche shapes. Mater.\ Res.\ Lett. 7:496-502.

\bibitem[Michel, 2001]{michel}
Michel, L., 2001. Fundamental concepts for the study of crystal symmetry. Phys.\ Rep. 341:265-336.

\bibitem[Miguel et al., 2001]{zapperi02}
Miguel, M.C., Vespignani, A., Zapperi, S., Weiss, J., Grasso, J.R., 2001. Intermittent dislocation flow in viscoplastic deformation. Nature 410:667-671.

\bibitem[Mumford et al, 2002]{dedekind1}
Mumford, D., Series, C., Wright, D., 2002. Indra's pearls. The visions of Felix Klein. Cambridge University Press.

\bibitem[Papanikolaou et al., 2017]{papanikolau}
Papanikolaou, S., Cui, Y., Ghoniem, N., 2017. Avalanches and Plastic Flow in Crystal Plasticity: An Overview. Mod.\ Sim.\ Mat.\ Sci.\  Eng.\ 26:013001.

\bibitem[Parry, 1998]{parry}
Parry, G.P., 1998. Low-dimensional lattice groups for the continuum mechanics of phase transitions in crystals. Arch.\ Rational Mech.\ Anal. 145:1-22.

\bibitem[Patriarca, 2019]{tesipatriarca}
Patriarca, C., 2019. Modular Energies for Crystal Elasto-Plasticity and Structural Phase Transformations. Master Thesis, Politecnico di Milano.

\bibitem[P\'erez-Reche, 2017]{pacoreview3}
P\'erez-Reche, F.-J., 2017. Modelling Avalanches in Martensites, in: Avalanches in Functional Materials and Geophysics. Understanding Complex Systems, Springer, 99-136.

\bibitem[P\'erez-Reche et al., 2016]{pacoreview2}
P\'erez-Reche, F-J., Triguero, C., Zanzotto, G., Truskinovsky, L., 2016. Origin of scale-free intermittency in structural first-order phase transitions. Phys.\ Rev.\ B 94:114102.

\bibitem[P\'erez-Reche et al., 2007]{pacoreview}
P\'erez-Reche, F.-J., Truskinovsky, L., Zanzotto, G., 2007. Training-induced criticality in martensites. Phys.\ Rev.\ Lett. 99:075501.

\bibitem[Pitteri, 1984]{pitterireconciliation}
Pitteri, M., 1984. Reconciliation of local and global symmetries of crystals. J.\ Elasticity 14:175-190.

\bibitem[Pitteri and Zanzotto, 2002]{PZbook}
Pitteri, M., Zanzotto, G., 2002. Continuum Models for Phase Transitions and Twinning in Crystals, Chapman \& Hall.

\bibitem[Qiu et al., 2019]{qiu19}
Qiu, D., Zhao, P., Shen, C., Lu, W., Zhang, D., Mrovec, M., Wang, Y., 2019. Predicting grain boundary structure and energy in BCC metals by integrated atomistic and phase-field modeling. Acta Mater. 164:799-809.

\bibitem[Quapp, 1989]{valleyfloor2}
Quapp, W., 1989. Gradient extremals and valley floor bifurcations on potential energy surfaces, Theor.\ Chim.\ Acta 75:447-460.

\bibitem[Quapp and Bofill, 2016]{saddleinflections5}
Quapp, W., Bofill, J.M., 2016. Reaction rates in a theory of mechanochemical pathways. J.\ Comput.\ Chem.\ 37: 2467-2478.

\bibitem[Rehbein and Carpenter, 2011]{saddleinflections4}
Rehbein, J., Carpenter, B.K., 2011. Do we fully understand what controls chemical selectivity? Phys.\ Chem.\ Chem.\ Phys. 13:20906-20922.

\bibitem[Salerno and Robbins, 2013]{salrob13}
Salerno, K.M., Robbins, M.O., 2013. Effect of inertia on sheared disordered solids: critical scaling of avalanches in two and three dimensions. Phys.\ Rev.\ E 88:062206.

\bibitem[Schoeneberg, 1975]{modularforms2}
Schoeneberg, B., 1975. Elliptic modular functions, Springer-Verlag, Berlin.

\bibitem[Shtyk et al., 2017]{electronsmonkeysaddle}
Shtyk, A., Goldstein, G., Chamon, C., 2017. Electrons at the monkey saddle: A multicritical Lifshitz point. Phys.\ Rev. B 95:035137.

\bibitem[Sills et al., 2018]{hardening}
Sills, R.B., Bertin, N., Aghaei, A., Cai, W., 2018. Dislocation Networks and the Microstructural Origin of Strain Hardening, Phys.\ Rev.\ Lett. 121:085501.

\bibitem[Song et al., 2013]{13james_nat}
Song, Y., Chen, X., Dabade, V., Shield, T.W., James, R. D., 2013. Enhanced reversibility and unusual microstructure of a phase-transforming material. Nature 502:85-88.

\bibitem[Sparks and Mass, 2018]{sparks1}
Sparks, G., Maass, R., 2018. Nontrivial scaling exponents of dislocation avalanches in microplasticity. Phys.\ Rev.\ Materials 2:120601.

\bibitem[Sparks and Mass, 2019]{sparks2}
Sparks, G., Maass, R., 2019. Effects of orientation and pre-deformation on velocity profiles of dislocation avalanches in gold microcrystals. Eur.\ Phys.\ J.\ B 92:15.

\bibitem[Stillwell, 2001]{dedekind3}
Stillwell, J., 2001. Modular Miracles. Am.\ Math.\ Monthly 108:70-76.

\bibitem[Stukowski, 2010]{ovito}
Stukowski, A., 2010. Visualization and analysis of atomistic simulation data with OVITO - the Open Visualization Tool. Model.\ Simul.\ Mater.\ Sci.\ Eng. 18:015012.

\bibitem[Sun et al., 2017]{prolifarationtwinning}
Sun, D., Ponga, M., Bhattacharya, K., Ortiz, M., 2017. Proliferation of twinning in hcp metals: Application to magnesium. J.\ Mech.\ Phys.\  Solids, 112:368-384.

\bibitem[Sun and Ruedenberg, 1993]{valleyfloor3}
Sun, J.Q., Ruedenberg, K., 1993. Gradient extremals and steepest descent lines on potential energy surfaces. J.\ Chem.\ Phys. 98:9707.

\bibitem[Tang, 2018]{tang18}
Tang, Y., 2018. Uncovering the inertia of dislocation motion and negative mechanical response in crystals. Sci.\ Rep. 8:140.

\bibitem[Terras, 2013]{poincarehalfplane1}
Terras, A., 2013. Harmonic Analysis on Symmetric Spaces - Euclidean Space, the Sphere, and the Poincar\'e Upper Half-Plane, Second Edition, Springer.

\bibitem[Truskinovsky and Vainchtein, 2004]{cottrell49fedelich92truvain042}
Truskinovsky, L., Vainchtein, A., 2004. The origin of nucleation peak in transformational plasticity. J.\ Mech.\ Phys.\ Solids 52:1421-1446.

\bibitem[Truskinovsky and Vainchtein, 2008]{truvain08}
Truskinovsky, L., Vainchtein, A., 2008. Dynamics of martensitic phase boundaries: discreteness, dissipation and inertia. Cont.\ Mech.\ Thermodyn. 20:97–122.

\bibitem[Uchic, 2009]{uchic092}
Uchic, M.D., Shade, P.A., Dimiduk, D.M, 2009. Plasticity of Micrometer-Scale Single Crystals in Compression. Ann.\ Rev.\ Mat.\ Res.\ 39:361-386.

\bibitem[Valtazanos and Ruedenberg, 1986]{saddleinflections3}
Valtazanos, P., Ruedenberg, K., 1986. Bifurcations and transition states. Theor.\ Chim.\ Acta 69:281-307.

\bibitem[Van Wijngaarden, 1953]{Jcoefficients}
Van Wijngaarden, A., 1953. On the coefficients of the modular invariant $J$. Proc.\ Kon.\ Nederl.\ Akad.\ Wetensch.\ A 16:389-400.

\bibitem[Vattr\'e and Denoual, 2016]{denoualJMPLplasticity}
Vattr\'e, A., Denoual, C., 2016. Polymorphism of iron at high pressure: A 3D phase-field model for displacive transitions with finite elastoplastic deformations. J.\ Mech.\ Phys.\ Solids 92:1-27.

\bibitem[Wales, 2003]{monkeybiomolecules}
Wales, D.J., 2003. Energy Landscapes: Applications to Clusters, Biomolecules and Glasses. Cambridge University Press.

\bibitem[Weiss et al., 2015]{differencesymmetry1}
Weiss, J., Ben Rhouma, W., Richeton, T., Dechanel, S., Louchet, F., Truskinovsky, L., 2015. From Mild to Wild Fluctuations in Crystal Plasticity. Phys.\ Rev.\ Lett. 114:105504.

\bibitem[Weiss, 2019]{weiss19}
Weiss, J., 2019. Ice: The paradigm of wild plasticity. Phil.\ Trans.\ Roy.\ Soc.\ A 377, https://doi.org/10.1098/rsta.2018.0260.

\bibitem[Xiang et al., 2017]{xiang}
Xiang, M., Cui, J., Yang, Y., Liao, Y., Wang, K., Chen, Y., Chen, J., 2017. Shock responses of nanoporous aluminum by molecular dynamics simulations. Int.\ J.\ Plast. 97:24-45.

\bibitem[Xie et al., 2019]{kan}
Xie, X., Kang, G., Kan, Q., Yu, C., Peng, Q., 2019. Phase field modeling to transformation induced plasticity in super-elastic NiTi shape memory alloy single crystal. Mod.\ Sim.\ Mat.\ Sci.\ Eng. 27:045001.

\bibitem[Ye et al., 2005]{dedekind5}
Ye, R-S., Zou, Y-R, Lu, J., 2005. Fractal Tiling with the Extended Modular Group. Lect.\ Notes in Comp.\ Sci., Springer, 3314:286-291.

\bibitem[Z\'ale\v z\'ak et al., 2017]{zalezak}
Z\'ale\v z\'ak, T., Svoboda, J., Dlouh\'y, A., 2017.  High temperature dislocation processes in precipitation hardened crystals investigated by a 3D discrete dislocation dynamics. Int.\ J.\ Plast. 97:1-23.

\bibitem[Zepeda-Ruiz et al., 2017]{plastMDbulatovnature}
Zepeda-Ruiz, L.A., Stukowski, A., Oppelstrup, T., Bulatov, V.V., 2017. Probing the limits of metal plasticity with molecular dynamics simulations. Nature 550:492.

\bibitem[Zhang et al., 2017]{differencesymmetry2}
Zhang, P., Salman, O.U., Zhang, J.-Y, Liu, G., Weiss, J., Truskinovsky, L., Sun, J., 2017. Taming intermittent plasticity at small scales. Acta Materialia 128:351-364.




\end{thebibliography}
\end{document}